\documentclass[preprint,12pt,authoryear]{elsarticle}

\usepackage{amssymb}
\usepackage[utf8]{inputenc}

\journal{Journal of Quantitative Spectroscopy \& Radiative Transfer}

\begin{document}

\begin{frontmatter}



\title{Beyond CCT: The spectral index system as a tool for the objective, 
quantitative characterization of lamps}


\author{D. Galad\'{\i}-Enr\'{\i}quez}
\ead{dgaladi@caha.es}
\address{Centro Astron\'omico Hispano-Alem\'an, Observatorio de Calar Alto, 
Sierra de los Filabres, ES-04550-G\'ergal (Almer\'{\i}a, Spain)}

\begin{abstract}

Correlated color temperature (CCT) is a semi-quantitative system that roughly 
describes the spectra of {lamps. This parameter gives 
the temperature (measured in kelvins)} of the black 
body that would show the hue more similar to that of the light emitted by the 
lamp.  Modern lamps for indoor and outdoor lighting display many spectral 
energy distributions, most of them extremely different to those of black bodies, 
what makes {CCT to be far from a 
perfect descriptor} from the physical point of view. The spectral index system 
presented in this work provides an accurate, objective, quantitative procedure to 
characterize the spectral properties of lamps, with just a
few numbers. The system is an adaptation to lighting technology of the 
classical procedures of multi-band astronomical photometry with wide and 
intermediate-band filters. We describe the basic concepts and we apply the system to
a representative set of lamps of many kinds. The results lead to
interesting, sometimes surprising conclusions.
The spectral index system is extremely easy to implement 
from the spectral data that are  routinely measured at laboratories. Thus, 
including this kind of computations in the standard protocols for the 
certification of lamps will be really straightforward, 
and will enrich the technical description of lighting devices.\newline
Sent to {\it JQSRT} 17 Oct 2017. Accepted 15 Dec 2017. Article available with DOI: 10.1016/j.jqsrt.2017.12.011 \newline
https://doi.org/10.1016/j.jqsrt.2017.12.011
\newline\newline\newline\newline\newline\newline
\noindent{\bf Highlights}
\begin{itemize}
\item A simple system for a physically meaningful, quantitative characterization of lamp spectra.
\item Spectral indices are straightforward to compute from the standard spectra currently obtained at any lab.
\item A natural link between lighting engineering and astrophysics, relevant for the study of artificial light at night.
\item A system potentially useful for industrial certification, legal regulation and biophysical studies.
\end{itemize}

\end{abstract}


\end{frontmatter}


\section{Introduction}
\label{intro}
Observational astronomy progressed for centuries having black bodies as its
almost only matter of study: the stars. And for millennia, the only
detector system in astronomy was the human eye, unaided or aided by
optical devices, what defined the sensitivity curve of human
sight as the only spectral band effectively available for the study of the universe.

The end of the \textsc{xix}$^{\mbox{\footnotesize{th}}}$ century brought the photographic revolution and,
with it, a different sensitivity curve that covered a slightly different spectral region,
biased towards bluer wavelengths. {Even though photographic emulsions are less
sensitive than the eye, this new technology allowed the study of much fainter celestial
objects, thanks to the possibility to accumulate light during very long exposure times.}

Approximately at the same time, spectroscopic techniques led to the discovery
of non-thermal emitters in astrophysical contexts: emission nebulae
whose light is made up mainly from narrow lines of ionized
atoms such as hydrogen, oxygen or sulfur. 

\begin{figure}
\resizebox{\textwidth}{!}{\includegraphics{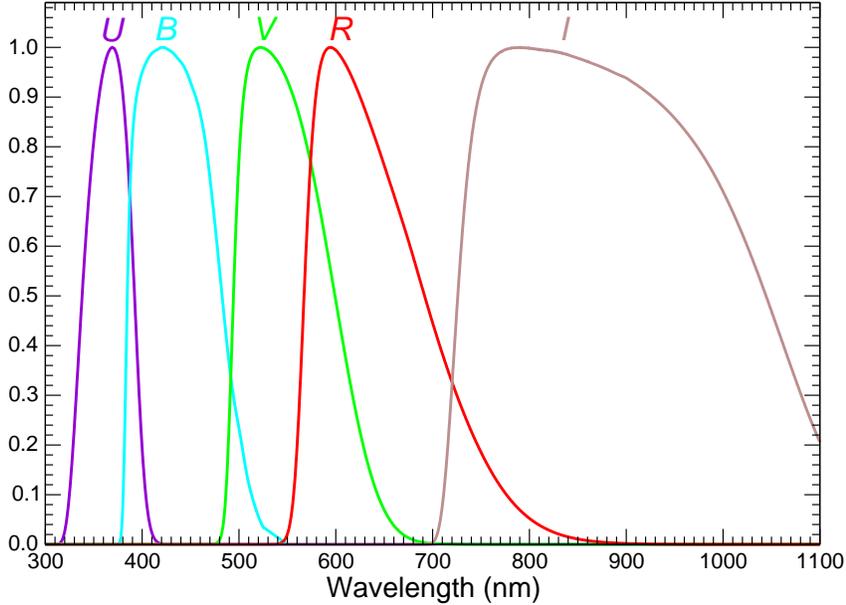}}
\caption{Normalized transmission curves for the five filters of the 
Johnson-Cousins system used in astronomical photometry. { More information in
Sect. \ref{basic} and Table \ref{Johnson_tab}.} See \cite{bes90}.}\label{Johnson_fig}
\end{figure}

More and more non-thermal
astrophysical sources have been discovered since then. 
Also, technological progress opened the whole electromagnetic spectrum
to  astrophysics, and many different bands have been defined, even inside
the optical window (that roughly covers from the near-UV to the near-IR). 
One of the most used photometric systems in observational astronomy
is the so called Johnson-Cousins, { based on a set of five filters (shown in  
Fig.~\ref{Johnson_fig}, and that will be commented later in Sect. \ref{basic} and Table \ref{Johnson_tab}), but many others exist.}

Interestingly enough, there exists a strong { analogy} between the evolution
of observational astronomy and that of lighting engineering, { that we have outlined in
Table \ref{analogy}}. Also this
field began with black bodies as the only working matter (combustion of
solids or gas, and incandescent lamps), and only one sensitivity curve
was considered at the beginning: The photopic (day-time) sensitivity curve
of the human eye. Somewhat later, the scotopic (darkness-adapted) 
sensitivity curve was added, and it was found to be much more sensitive,
and biased towards the blue. Later on, new light sources have appeared,
that are not thermal emitters, such as discharge lamps and light-emitting 
diodes (LEDs). 

\begin{table*}
\caption{A comparison of observational astronomy and lighting technology evolution.}\label{analogy}
\begin{center}
\begin{tabular}[]{p{43mm}|p{43mm}|p{43mm}}
\hline
    & {\centering {\bf Observational \newline astronomy}}  & {\centering {\bf Lighting technology}} \\
\hline
{\raggedright {\bf Original spectra}}     & Black bodies (stars) & Black bodies (incandescent lamps) \\
\hline
{\raggedright {\bf Original band}}        & Human eye sensitivity curve  & Human eye sensitivity  curve (photopic) \\
\hline
{\raggedright {\bf First additional band}}& Photography (biased towards blue) & Scotopic sensitivity curve (biased towards blue) \\
\hline
{\raggedright {\bf Additional spectra}}   & Non-thermal emitters: emission nebulae; synchrotron radiation; etc. Red-shifted spectra. 
                                   & Non-thermal emitters: discharge lamps; light emitting diodes. \\
                                   \hline
{\raggedright {\bf Later additional \newline bands}}& Many bands inside and outside the visible spectrum. For instance, Johnson-Cousins
                                     photometric system $UBVRI$. &
                                     Bands linked to visual and non-visual pigments: {\it LC}, {\it MC},
                                     $R$, $Z$, {\it CS}. Non-human sensitivity curves. \\
\hline
\end{tabular}
\end{center}
\end{table*}

Huge advancements have happened in recent times, 
in the research of photo-sensitive pigments in humans and in other species
both animal and vegetal, what implied characterizing many spectral
sensitivity curves that complement the traditional ones. In this
context of non-thermal emitters and multiplicity of spectral bands,
analog to the evolution experienced in astronomy, it arises the need
to re-think those concepts used in lighting engineering that
are based upon the properties of human vision and of thermal
light sources.

{ Correlated Color Temperature (CCT) is a way to link human perception of the hue
of lamps to the thermodynamic temperature of black bodies. 
The current official definition can be found at \cite{cie04}.
For non-black-bodies, CCT 
lacks rigorous physical meaning and it provides just a perceptual indication
of the hue of the light. Significant differences in the perceived
hue are admitted, even for sources having the same CCT.} 
In a multi-band (even non-human-band) and non-thermal context, CCT loses most of its meaning,
even in spite the efforts to bring this parameter to the limit of
maximum {\it numerical} accuracy as in \cite{cha16}. 
We have to ask ourselves whether better methods do exist,
to characterize the spectral properties of lamps. From the
qualitative and unsatisfactory, one-number CCT descriptor, to the heavy
power of giving the whole spectrum in high resolution as suggested
by \cite{luc14}, there has to be some
middle point that allows us to work with just a few numbers, with
univocal and clear physical meanings, that should even make it possible to 
perform meaningful calculations (something completely out of place with CCT). 

We explore the promising prospects that arise from the adaptation of 
some of the techniques developed in multi-band
astronomical photometry, to lighting engineering. In particular we 
will work on the so-called color index system,
that we propose to translate into a format suited to the description of lamps
under the name of \textit{spectral index system}: Converging solutions for 
two fields of study that have followed parallel trajectories during the 
last two centuries.

\section{The spectral index system}

{ Fortunately, the two worlds that converge into this scheme, astronomical 
photometry and lighting engineering, follow traditions fully compatible in what refers to
the conventions used to describe spectral energy distributions and filters,
what allows a soft join that may benefit both fields.}

\subsection{Basic concepts: spectrum and filter}
\label{basic}
Let us start commenting 
the basic ingredients needed to compute spectral indices: The spectral
energy distribution of the light source, and the transmission curves of 
filters, equivalent in many senses to the spectral sensitivity curves of photopigments. Both communities
prefer working in terms of wavelength ($\lambda$) rather than frequency ($\nu$), and
the standard way of characterizing filters and sensitivity curves is
normalizing them in such a way that the maximum is set to unity. In this paper,
all wavelengths are measured in nanometers (nm). Spectra, also called spectral
energy distributions (SED), are noted as $E(\lambda)$. Filters (or sensitivity
curves) are noted as $F(\lambda)$. 

Table~\ref{Johnson_tab} displays some descriptors for a set of filters of interest.
Among these filters, there are the five of the astronomical 
photometric system Johnson-Cousins (see \cite{bes90} for more details). 
{ To avoid confusions that may arise from the coincidence of symbols, in this work
we will label Johnson-Cousins filters as 
$U_{\mbox{\footnotesize J}}$, 
$B_{\mbox{\footnotesize J}}$, 
$V_{\mbox{\footnotesize J}}$, 
$R_{\mbox{\footnotesize C}}$, 
$I_{\mbox{\footnotesize C}}$.
Table \ref{Johnson_tab} includes, also, the photopic $V$ and scotopic $V'$ sensitivity curves
of human vision (see \cite{vos78} for $V$, and \cite{wys82} or \cite{sch04} for $V'$), 
and the spectral sensitivity curves of five human 
photosensitive pigments given by \cite{cie15}: cyanopic ($SC$), melanopic ($Z$), rhodopic ($R$), 
chloropic ($MC$) and erythropic ($LC$), mentioned in Sect.~\ref{curves}. }
The filter descriptors are effective wavelength $\lambda_{\mbox{eff}}$ and filter
width $\Delta\lambda$, defined as follows:

\begin{equation}
\label{descriptors}
\lambda_{\mbox{eff}}=\frac{\int_0^\infty \lambda F(\lambda) \mbox{d}\lambda}
{\int_0^\infty F(\lambda) \mbox{d}\lambda}; \;\;\;
\Delta\lambda=\frac{1}{F_{\mbox{max}}}\int_0^\infty F(\lambda) \mbox{d}\lambda
\end{equation}

Where $F_{\mbox{max}}$ stands for the maximum value of the filter curve $F(\lambda)$,
that normally will be equal to unity. { Note that, in general, $\lambda_{\mbox{eff}}$ 
is not equal to the wavelength at which the curve $F(\lambda)$ reaches its maximum, although the
two values should be very similar for filter functions symmetric in shape.}

\begin{table*}
\caption{Effective wavelengths and filter widths (see definitions in Eq. 
\ref{descriptors}) for several sensitivity curves: Those of the astronomical
photometric system Johnson-Cousins, photopic and scotopic curves $V$ and $V'$, and several 
human photopigments.}\label{Johnson_tab}
\begin{center}
\begin{tabular}[c]{l|ccccc}
\hline
 & \multicolumn{5}{c}{Johnson-Cousins} \\
 
Filter: & $U_{\mbox{\footnotesize J}}$ 
        & $B_{\mbox{\footnotesize J}}$ 
        & $V_{\mbox{\footnotesize J}}$ 
        & $R_{\mbox{\footnotesize C}}$ 
        & $I_{\mbox{\footnotesize C}}$  \\
\hline
$\lambda_{\mbox{eff}}$ (nm): & 365.3 & 438.2 & 552.4 & 645.2 & 885.9 \\ 
$\Delta\lambda$ (nm):        &  53.1 &  98.0 & 104.8 & 129.9 & 300.6 \\ 
\hline
\hline
 & \multicolumn{5}{c}{Human vision curves} \\
Filter: &  & $V'$ & $V$ &  &  \\ 
\hline
$\lambda_{\mbox{eff}}$ (nm): &  & 502.4 & 559.4 &  &   \\
$\Delta\lambda$ (nm):        &  &  97.1 & 107.4 &  &   \\
\hline
\hline
 & \multicolumn{5}{c}{Photopigments} \\
Filter: & {\it SC} & $Z$ & $R$ & {\it MC} & {\it LC} \\
\hline
$\lambda_{\mbox{eff}}$ (nm): & 452.8 & 496.5 & 512.8 & 542.5 & 566.9  \\
$\Delta\lambda$ (nm):        &  51.4 &  83.1 &  96.9 & 110.8 & 118.6  \\
\hline
\end{tabular}
\end{center}
\end{table*}

As we will show later, the units of $E(\lambda)$ are not particularly relevant,
as long as they are expressed in terms of physical energy (not photon counts),
and the function is well calibrated, not affected by instrumental biases nor
other spectral filtering. These conditions rule for the material
routinely produced in lighting engineering for lamp certification purposes.
The spectrograph output at the laboratory may be expressed in terms
of W/nm, $\mu$W/(cm$^2$ nm) at some standard distance, W/(m$^2$ sr nm), etc. 
As said, any of these will work perfectly when fed into our formalism. 

From the filtered spectrum, $F(\lambda)E(\lambda)$, through integration, we
get the \textit{integrated flux}, $\Phi_{E,F}$:

\begin{equation}
\label{int_flux}
\Phi_{E,F}=\int_{0}^{\infty} F(\lambda)E(\lambda)\mathrm{d}\lambda
\end{equation}

A particular case of filter is set by the absence of any filter at all or,
in other words, $F(\lambda)=1\;\forall\lambda$. We refer to this non-filter
as the \textit{bolometric filter}, and it leads to the \textit{bolometric flux}:

\begin{equation}
\label{bol_flux}
\Phi_{E,\mbox{\footnotesize{bol}}}=\int_{0}^{\infty} E(\lambda)\mathrm{d}\lambda
\end{equation}

\subsection{Definition of the spectral index}
\label{definition}
The \textit{relative integrated flux} among two filters $F_1$ and $F_2$,
for a given spectrum $E$, is given by the quotient of integrated fluxes:

\begin{equation}
\label{rel_flux}
Q_{1,2}(E)=\frac{\Phi_{E,F1}}{\Phi_{E,F2}}=
\frac{\int_{0}^{\infty}F_1(\lambda)E(\lambda)\mathrm{d}\lambda}
     {\int_{0}^{\infty}F_2(\lambda)E(\lambda)\mathrm{d}\lambda}
\end{equation}

The self-normalization implicit in Eq.~\ref{rel_flux} makes it evident
that the specific units in which $E(\lambda)$ is expressed are not relevant.

Finally, the spectral index of spectrum $E$ for the pair of filters $F_1$ and
$F_2$ is defined this way:

\begin{eqnarray}
\label{spectral_index}
C_{1,2}(E) & = & -2.5\log_{10}Q_{1,2}(E)=-2.5\log_{10}\frac{\Phi_{E,F1}}{\Phi_{E,F2}}= \nonumber \\
           & = & -2.5\log_{10}\int_{0}^{\infty}F_1(\lambda)E(\lambda)\mathrm{d}\lambda \\
           &   & +2.5\log_{10}\int_{0}^{\infty}F_2(\lambda)E(\lambda)\mathrm{d}\lambda \nonumber
\end{eqnarray}

The quantity $-2.5\log_{10}\Phi_{E,Fi}; i=1,2$, that appears in Eq.~\ref{spectral_index},
is named \textit{instrumental magnitude of E in filter $F_i$} and it may be also represented
as $m_{Fi}(E)$. Thus, we can say that a spectral index is a difference of instrumental
magnitudes:

\begin{equation}
\label{instrumag}
C_{1,2}(E)=m_{F1}(E)-m_{F2}(E)
\end{equation}

{ Magnitudes as a system to evaluate the apparent brightness of stars have been in use 
in astronomy for more than two thousand years, in a tradition that can be traced
back at least to the time of Hipparchus of Nicaea (around year 150 B.C., see for instance 
Ch. 4 in \cite{north95}). This system is rooted into the peculiarities of human vision,
what justifies its logarithmic nature, the instrumental magnitude being simply 
$-2.5$ times the decimal logarithm of the integrated flux.
It is very important to realize that with the negative sign introduced in the definition
of instrumental magnitude, a larger integrated flux implies a lower numerical value
for the associated instrumental magnitude. Number 2.5 fixes the scale in such a way that
a difference of 5 magnitudes implies a factor 100 in integrated flux, and it was implicitly
introduced by Hipparchus when he established that the fainter stars seen with the naked eye
have magnitude equal to six, while the brighter ones have magnitude equal to one. 
There is a second and unexepected link of the magnitude scale with 
human vision: As we will see later (Sect. \ref{generic_generic}), 
the sensitivity contrast of human vision among photopic and scotopic conditions
amounts almost exactly to one magnitude.}

For the bolometric filter (i.e., in absence of filter) we get the bolometric instrumental magnitude,
that is a measure of the total emission of the lamp across the whole spectrum:

\begin{equation}
\label{bolomag}
m_{\mbox{\footnotesize{bol}}}(E) = -2.5\log_{10}\int_{0}^{\infty}E(\lambda)\mathrm{d}\lambda
\end{equation}

An important convention, linked to the definition expressed in Eq.~\ref{spectral_index},
is the need to sort the two filters, $F_1$ and $F_2$, in such a way that the
first one is always the bluest of the pair. This convention leads to spectral indices
with larger numerical values for redder sources, and smaller (even negative)
values for bluer spectra. We will refer to this convention as the {\it bluer first} rule.

The election of decimal logarithms, the negative sign and the 2.5 coefficient may seem
arbitrary, but they are directly drawn from the metrological system already in use in
astronomical photometry, that already is widely spread  { to measure the brightness of astronomical
sources of radiation like the stars, but also to describe} natural
and artificial sky brightness. The effects of artificial light at night on sky brightness 
is commonly measured in the stellar magnitude scale, and it leads in a very natural
way to the evaluation of sky color in the same system (see for instance \cite{sdm17}). Thus,
using the same methodology for the description of lamp spectra and its effects,
will establish an interesting bridge between lighting engineering on the one side, and astronomy and 
the study of artificial light at night on the other side.

When the relative flux (Eq.~\ref{rel_flux}) is of interest on itself, it can be 
retrieved from the corresponding spectral index in  a straightforward way, inverting the 
definition (Eq.~\ref{spectral_index}):

\begin{equation}
\label{C_t_Q}
Q_{1,2}(E)=\frac{\Phi_{E,F1}}{\Phi_{E,F2}}=10^{-C_{1,2}(E)/2.5}
\end{equation}

\subsection{Sensitivity curves used}
\label{curves}

\begin{figure}
\resizebox{\textwidth}{!}{\includegraphics{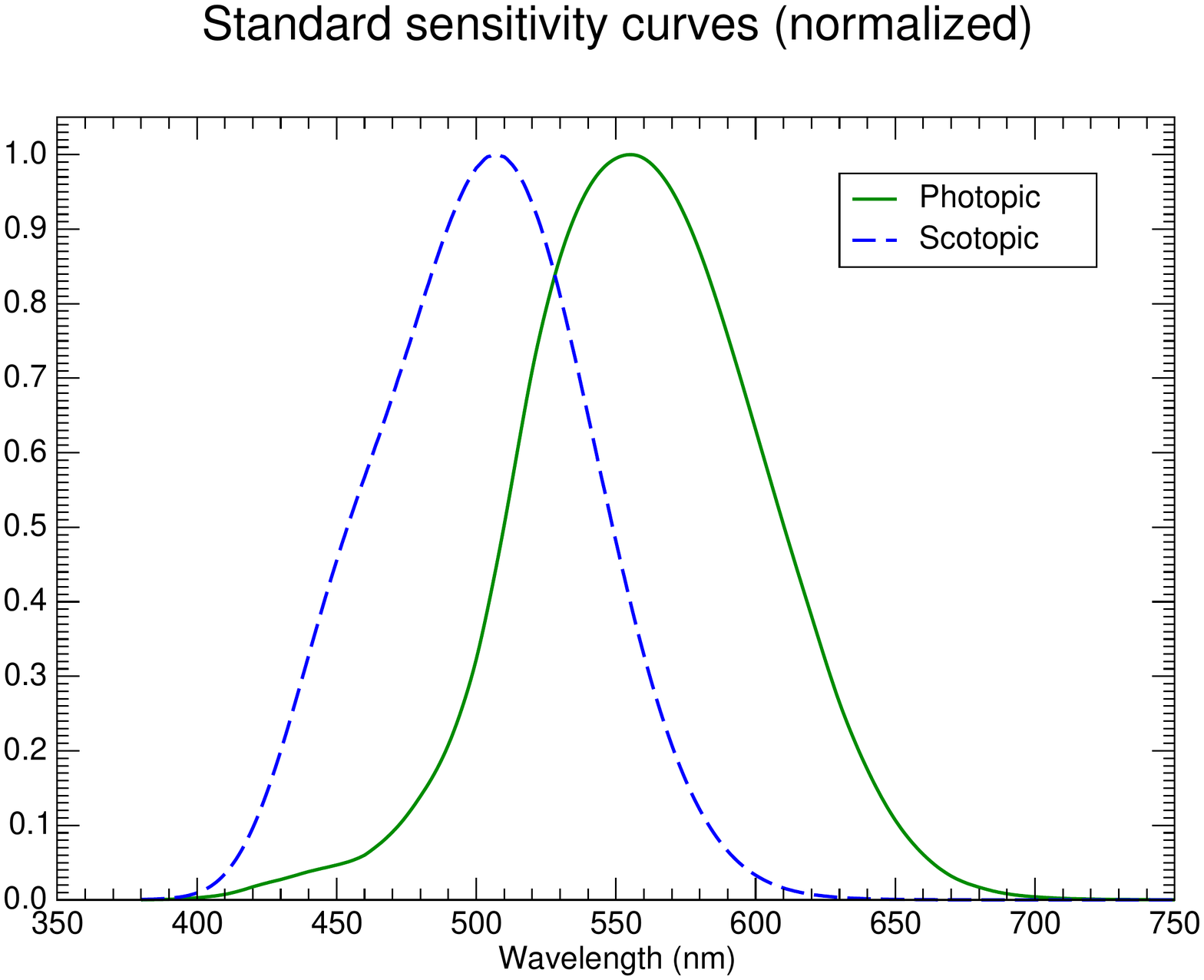}}
\caption{Normalized human eye standard sensitivity curves under photopic ($V$, solid line)
and scotopic ($V'$, dashed) conditions. See \cite{vos78}, \cite{wys82} and \cite{sch04}.}\label{VandVprime}
\end{figure}

The simple formalism sketched in Sect.~\ref{definition} is absolutely general.
In order to apply it we need, of course, some specific spectrum $E(\lambda)$ but,
obviously, two spectral bands have to be selected and defined. Let us
consider which filters or spectral sensitivity curves may be of interest 
for lighting technology. No doubt, the number of such bands may be very high.
Among them we have to include the standard sensitivity curves of human vision,
photopic $V(\lambda)$ and scotopic $V'(\lambda)$, described in Fig.~\ref{VandVprime} 
and in Table~\ref{Johnson_tab}. The similarities between the astronomical
Johnson $V_{\mbox{\footnotesize J}}$ band and the photopic $V$ function are very obvious and they are not
casual, since the astronomical filter was specifically designed to have a match
as good as possible with the sensitivity of human sight. 

\begin{figure}
\resizebox{\textwidth}{!}{\includegraphics{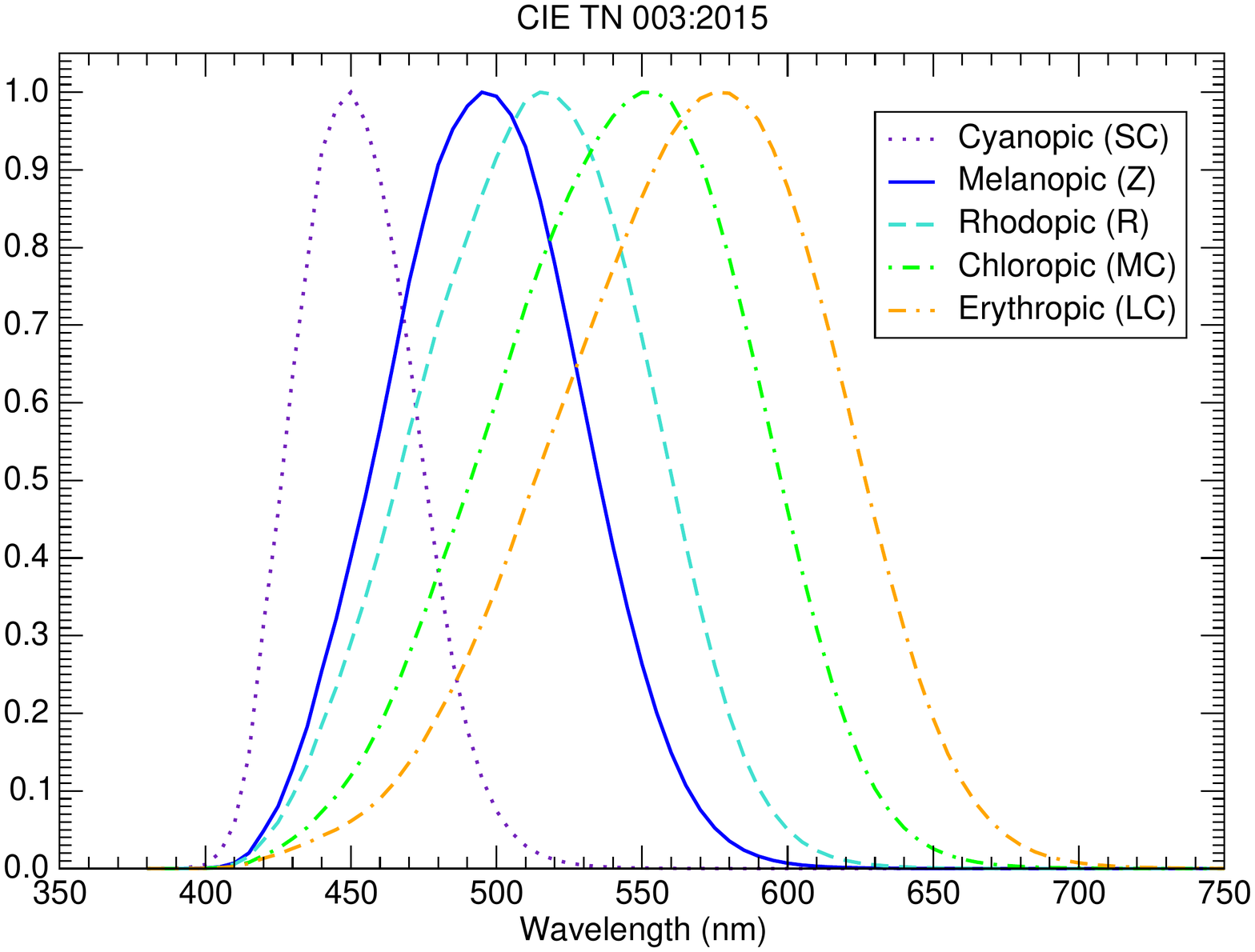}}
\caption{Normalized sensitivity curves for the five human photo-receptors, after
applying the pre-receptoral transmittance function. See Annex A to \cite{cie15}.}\label{CIEcurves}
\end{figure}

\begin{figure}
\resizebox{\textwidth}{!}{\includegraphics{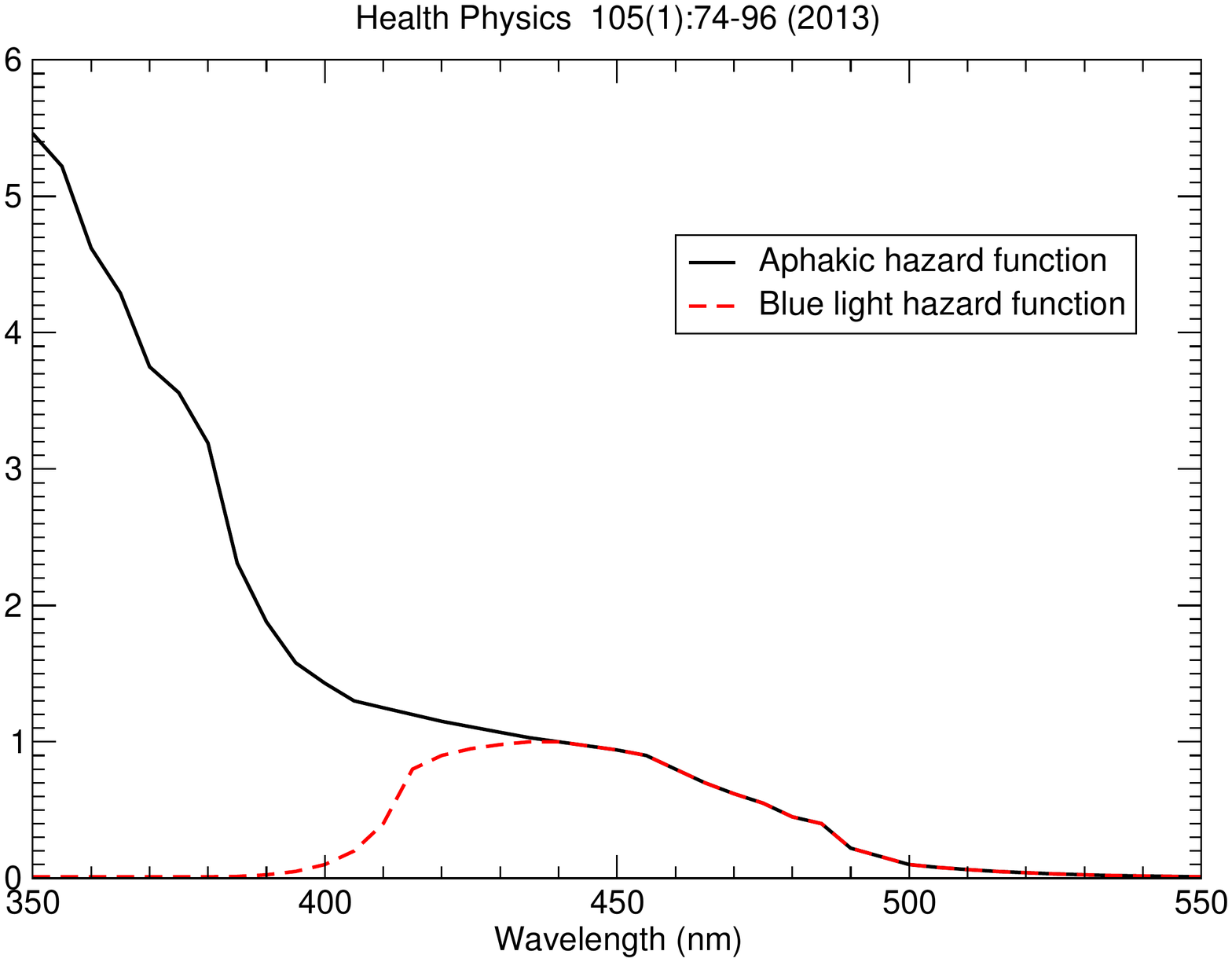}}
\caption{Action curves named blue-hazard function ($B$, dashed line) and aphakic hazard function
($A$, solid line), from \cite{icnirp13}. The descriptors of $B$ function (Eqs.~\ref{descriptors})
are $\lambda_{\mbox {eff}}=446.2$ nm, $\Delta\lambda=69.0$ nm, not too far from the astronomical
Johnson $B_{\mbox{\footnotesize J}}$ band (Table~\ref{Johnson_tab}, Fig.~\ref{Johnson_fig}). (For interpretation of the references to color in this figure legend, the reader is referred to the web version of this article.)}\label{blue_hazard}
\end{figure}

\cite{cie15} defines the spectral sensitivity curves of five human 
photosensitive pigments, together with a common pre-receptoral
transmittance curve that evaluates the spectral absorption of the 
tissues placed in front of the receptors themselves. From the table
in Annex A of \cite{cie15} we take five sensitivity curves 
named cyanopic $(SC)$, melanopic $(Z)$, rhodopic $(R)$, 
chloropic $(MC)$ and erythropic $(LC)$. Applying to all of them the associated 
pre-receptoral transmittance curve $\tau(\lambda)$, and later normalizing the 
resulting curves to max = 1, we get the sensitivity curves described in
Table~\ref{Johnson_tab} and Fig.~\ref{CIEcurves}.

The International Commission on Non-Ionizing Radiation Protection
defines in \cite{icnirp13} the blue light hazard function $(B)$
and the aphakic hazard function $(A)$ (see Fig.~\ref{blue_hazard}). For the reasons
described in the source publication, { and commented in Sect. \ref{discussion}
in the discussion of Fig. \ref{bluehazardCCT}, curve $A$ cannot be normalized
to max = 1, what has to be kept in mind when interpreting results
involving this filter.} 

Several interesting, and very simple bandpasses can be defined as
step functions, specially filters transparent to only short
wavelengths. We label such filters as $L_{\lambda 0}$, where 
$\lambda_0$ is such that:

\begin{equation}
\label{lowpass}
L_{\lambda 0} = \left\{ \begin{array}{ll}
                 1 & \forall \; \lambda \leq \lambda_0 \\
                 0 & \mbox{otherwise}
                         \end{array}
                \right.
\end{equation}

We will refer to such filters as \textit{lowpass}-$\lambda$, because we are
working in the wavelength space and those filters block high values
of $\lambda$, leaving low values unaltered. 
Of course, the same physical filter (maybe made from
glass) would be described as a \textit{highpass} if we were working 
in the frequency space. 

{
In a similar way, highpass-$\lambda$ filters may be defined as:

\begin{equation}
\label{highpass}
H_{\lambda 0} = \left\{ \begin{array}{ll}
                 0 & \forall \; \lambda \leq \lambda_0 \\
                 1 & \mbox{otherwise}
                         \end{array}
                \right.
\end{equation}

The product of a highpass by a lowpass specifies a window filter. Thus, filter
$H_x \times L_y$ would be transparent to wavelengths between $x$ and $y$ nm.

Lowpass-$\lambda$, highpass-$\lambda$ and window filters are important 
because there are already several
regulations and recommendations establishing spectral restrictions
on lamps, on the basis of the quantity of radiation emitted 
below, above or between certain specific wavelengths. As an example, 
Table \ref{legal} (that will be commented in Sect. \ref{useful})
summarizes some of these already existing
specifications, translating them into the language of  
spectral indices. 
}

\section{Some useful specific indices}
\label{useful}
Now we get closer to the specific application of the formalism. In order to do 
that, in this section we review several pairs of filters that lead
to spectral indices meaningful for the description of lamp spectra. 
First we consider filter pairs that include the bolometric filter 
to derive \textit{bolometric indices}. In a second step we review
indices implying lowpass-$\lambda$ filters. Finally, we describe
some generic indices made up from filter pairs of any kind.

\subsection{Bolometric indices}
\label{bolom_indices}
We talk about \textit{bolometric indices} when dealing with 
spectral indices that include the bolometric filter as one of 
the two sensitivity curves involved in the calculation (Eq.~\ref{spectral_index}). 
The {\it bluer first} convention linked to the definition of spectral indices 
(Sect.~\ref{definition}) requires that the first filter of the 
pair has to be the bluer one. The bolometric filter, being described
by a constant function equal to unity for all values of 
$\lambda$, is in fact an infinitely red filter. Thus, it is always the redder of
any pair and has to be introduced in the formulae as $F_2$. 
Representing the first filter as $F$, whatever it is, and the bolometric
one (the second) as bol, we have the definition of bolometric
index of spectrum $E$ for filter $F$:

\begin{equation}
\label{boloindex}
C_{F,\mbox{\footnotesize{bol}}}(E) = m_{F}(E) - m_{\mbox{\footnotesize{bol}}}(E)
\end{equation}

It is evident that under these conditions we will always have more radiation
in the second filter (the bolometric one means absence of any filter),
what will make $m_{\mbox{\footnotesize{bol}}}(E)$ smaller (i.e., 'brighter',
due to the negative sign in the definition of instrumental magnitudes). As a 
conclusion, all bolometric indices will always be positive. They provide a 
way to evaluate the portion of energy emitted by a lamp in the band covered
by filter $F$, compared to the total emission over the whole spectrum.

\begin{figure}
\resizebox{\textwidth}{!}{\includegraphics{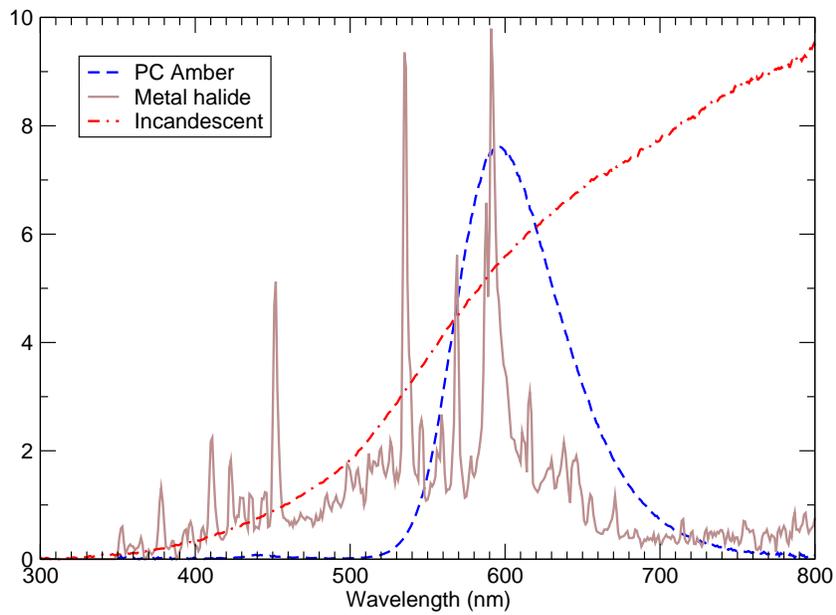}}
\caption{Spectra of three lamps used as examples for the application
of the spectral index system. A PC amber lamp (number 39 in the ancillary database \cite{gal17}),
a discharge metal halide lamp (number 25) and a classical incandescent bulb (number 29). \label{sample_lamps}
}
\end{figure}

It is time to start applying the formalism to specific lamps.
Let us take as first examples the spectra displayed in Fig.~\ref{sample_lamps}, corresponding to
a PC amber LED lamp, a metal halide lamp, and a standard
incandescent bulb. If we take the photopic sensitivity curve
$V$ as $F_1$, the corresponding bolometric index will measure the amount of 
useful radiation (from a photopic point of view) emitted by the lamps, compared
to their total emission. We get the results:

\begin{center}
\begin{tabular}{lrr}
\hline
\multicolumn{1}{c}{$E$} & 
\multicolumn{1}{c}{$C_{V,\mbox{\footnotesize{bol}}}(E)$} &
\multicolumn{1}{c}{$Q_{V,\mbox{\footnotesize{bol}}}(E)$} \\
\hline
PC amber     & 0.721 & 0.515 \\
Metal halide & 0.819 & 0.470 \\
Incandescent & 2.096 & 0.145 \\
\hline
\end{tabular}
\end{center}

These figures indicate a higher visual efficacy of the PC amber lamp, if the 
photopic emissions are compared to the total amount of radiation. In fact,
the visible light emitted by the PC amber lamp is larger than 50\% of the total, 
because its index is lower than $0.753 = -2.5 \log_{10}(0.5)$. The lower
photopic efficacy corresponds in this case to the incandescent bulb:
its very red (high value) index is due to its large amount of infra-red emission,
and it has to be taken into account that the experimental spectrum used
is truncated at certain infra-red wavelength, so the true value for its 
$C_{V,\mbox{\footnotesize{bol}}}$ index should be even larger, meaning 
in this case an even lower light efficacy. 

The maximum efficacy in the bolometric index for an arbitrary filter would correspond to 
a lamp emitting monochromatic light at the wavelength in which the first filter
curve reaches unity, and in this case the bolometric index would be 
equal to zero. The closer the bolometric index gets to zero, the higher the
lamp efficacy for the filter used to perform the computations.

\begin{table*}
\caption{Some existing regulations whose requirements may be expressed in terms
of spectral indices.}\label{legal}
\begin{center}
\begin{tabular}[]{p{40mm}|p{27mm}|p{63mm}}
\hline
{\centering {\bf Location}}  & {\centering {\bf Reference}} & {\centering {\bf Requirements}} \\
\hline
Andalusia (Spain)     & \cite{boja10} \newline Article 13.a 
& In general: \newline 
$L_{525}-{\mbox {bol}} > 0.753$  \newline
At protected areas: \newline
$L_{440}-{\mbox {bol}} > 2.060$ for non-LED \newline
$L_{500}-{\mbox {bol}} > 2.060$ for LED \\
\hline
Canary Islands \newline (Spain) & \cite{iac17} \newline Section G.8   
& IAC amber LED definition: \newline
$L_{500}-{\mbox {bol}} > 4.560$ \newline
IAC warm LED definition: \newline
$L_{500}-V > 1.505 $ \newline
IAC super-warm LED definition: \newline
$L_{500}-V > 2.060$ \\
\hline
Antofagasta, Atacama and Coquimbo (Chile) & \cite{chile13} \newline Article 7 
&  Simultaneously required: \newline
$H_{300}\times L_{379} - H_{380}\times L_{780} > 2.060 $ \newline
$H_{380}\times L_{499} - H_{380}\times L_{780} > 2.060 $ \newline
$H_{380}\times L_{780} - H_{781}\times L_{1000} < -0.753 $ \\
\hline
\end{tabular}
\end{center}
\end{table*}

An interesting case is that posed by low- and highpass-$\lambda$ bolometric indices. 
There are already some regulations and recommendations on lamp spectra
that establish a certain limit for the fraction of the total radiation 
emitted below or over a specific wavelength, or inside certain intervals.
{ 
Table \ref{legal} shows several examples. Specifically the lowpass -- bolometric
criteria is included in the 
Andalusian regulation (\cite{boja10}), and in the Instituto de Astrof\'{\i}sica
de Canarias (IAC) amber LED definition from
\cite{iac17}. For instance, in one case it is 
required that the total emission below 500 nm has to be less than 
15\% of the total (Andalusian requirement for LEDs at protected areas). 
That means that the $L_{500}-$bol index has to be
{\it larger} than 2.060. In the case of the same three previous lamps, we get:

\begin{center}
\begin{tabular}{lrr}
\hline
\multicolumn{1}{c}{$E$} & 
\multicolumn{1}{c}{$C_{L500,\mbox{\footnotesize{bol}}}(E)$} &
\multicolumn{1}{c}{$Q_{L500,\mbox{\footnotesize{bol}}}(E)$} \\
\hline
PC amber      & 5.662 & 0.005 \\
Metal halide  & 1.652 & 0.218 \\
Incandescent  & 3.618 & 0.036 \\
\hline
\end{tabular}
\end{center}

In this case, the PC amber lamp would widely fulfill the requirement, but this
specific metal halide lamp would not. It has to be noted that also 
the incandescent lamp taken as example fulfills this limit. 
This PC amber lamp qualifies, too, as ''IAC amber LED'', because its 
index is larger than 4.560 (see Table \ref{legal}). 
}

However, bolometric criteria can be criticized, because they measure 
efficacy comparing a certain band to the total amount of emission,
including even non-visible wavelengths. For instance,
the above-mentioned criterium based on a lowpass-$\lambda$ bolometric
index, very clearly favours lamps with strong infra-red 
emissions. If going beyond a certain value
of $C_{L500,\mbox{\footnotesize{bol}}}$ is required, this can be 
achieved not only by reducing the amount of light at the blue side,
but also by increasing the wasteful infra-red radiation at the red side. 
This is the main reason why incandescent lamps display such large
values for indices of this kind, as can be seen in the data accompanying
this article, \cite{gal17}.

\subsection{Generic indices for pass-$\lambda$ filters}
\label{generic_lowpass}
That caveat can be easily overcome just setting the right, 
non-bolometric band, as second filter for the calculations. 
An obvious enough option would be using the photopic curve
$V$ as a reference. { This curve has its effective wavelength around
$\lambda=560$ nm. The central wavelength of any lowpass-$\lambda$ filter $L_{x}$ 
is placed exactly at $x$/2}. Given that normally the aim
is to limit the amount of emission in blue bands,
most often { $x/2$ will be smaller than 560 nm, and} 
the step filter will be the first {\it (bluer first)}, and the photopic 
curve will act as the second {\it (redder)} band for spectral index calculation. 
The resulting index $L_{x}-V$ would describe the quantity of
energy emitted in the blue, below $\lambda = x$ nm, compared
to the amount of photopically efficient light, what seems
a fair and meaningful comparison. 

{ The recent update of the regulations at Canary Islands 
(see \cite{iac17}) includes several specifications that may be
easily translated into our formalism using exactly this index. 
They appear in Table
\ref{legal}, and they refer to the definitions of ''IAC warm LED''
and ''IAC super-warm LED'', that require the index 
$L_{500}-V$ to reach at least the values 1.505 (warm) or
2.060 (super-warm). Going back to the 
same three lamps that we are using as an example, for 
$L_{500}-V$ we get the results:

\begin{center}
\begin{tabular}{lrr}
\hline
\multicolumn{1}{c}{$E$} & 
\multicolumn{1}{c}{$C_{L500,V}(E)$} &
\multicolumn{1}{c}{$Q_{L500,V}(E)$} \\
\hline
PC amber     & 4.491 & 0.016 \\
Metal halide & 0.833 & 0.464 \\
Incandescent & 1.522 & 0.246 \\
\hline
\end{tabular}
\end{center}

Not surprisingly, the PC amber LED overruns both
criteria, but also the incandescent lamp would fit in the 
''warm'' box (although these IAC definitions are intended only for LEDs). 
Now we get a significantly bluer (lower) value
of the index for the incandescent bulb, compared to the bolometric result,
because for index $L_{500}-V$ all infra-red 
emissions are kept out of the calculation. Something similar,
to a lesser extent, happens to the metal halide lamp.}

\subsection{Totally generic indices}
\label{generic_generic}
The flexibility of the spectral index system arises from its 
totally generic character, allowing to select any pair of 
spectral bands of interest. 

{
As an illustration of the general scheme, let us translate into
the spectral index formalism a classical photometric parameter, the 
so-called ''scotopic to photopic ratio'', or $S/P$. Traditionally, this
ratio is computed from the lamp spectrum $E(\lambda)$, filtered
trough the two standard sensitivity curves of human vision displayed 
in Fig. \ref{CIEcurves}. The scotopic $V'(\lambda)$ and photopic
$V(\lambda)$ functions are normalized to maximum equal to unity, and
they can be introduced into Eq. \ref{spectral_index} as $F_1$ and $F_2$
{\it (bluer first)} to compute the ratio 
$Q_{V',V}(E)=\frac{\Phi_{E,V'}}{\Phi_{E,V}}$ and, from there,
the scotopic -- photopic spectral index $C_{V',V}(E)$ for the lamp.
Applying this to the same three examples, we get:

\begin{center}
\begin{tabular}{lrr}
\hline
\multicolumn{1}{c}{$E$} & 
\multicolumn{1}{c}{$C_{V',V}(E)$} &
\multicolumn{1}{c}{$Q_{V',V}(E)$} \\
\hline
PC amber      &  2.032 & 0.154 \\
Metal halide  &  0.521 & 0.619 \\
Incandescent  &  0.796 & 0.480 \\
\hline
\end{tabular}
\end{center}

In this scheme, a null value would mean ''same energy in both bands''.
The first lamp (PC amber) has an index value close to 2, meaning that
the blue filter (scotopic) is receiving approximately 15\% of the energy
that goes through the red (photopic) one. The value for the incandescent 
lamp is close to the 50\% energy ratio (photopically efficient energy
doubles the amount of scotopically efficient flux). 

But normally the $S/P$ ratio is computed not from the normalized
$V'$ and $V$ functions, but from the scaled versions of these functions,
including the well-known scaling factors 1700 lm/W for the scotopic curve, 
and 683 lm/W for the photopic:

\begin{equation}
\label{scaled}
K'(\lambda) = 1700 \; V'(\lambda);  \;\;\;\;
K(\lambda) = 683 \; V(\lambda)
\end{equation}

This way we move from the energy domain into the human perceptual domain, going 
further from the physical input and closer to the action exerted by this input. 
The logarithmic nature of the spectral index allows a straightforward transformation
from the normalized index $C_{V',V}(E)$ to the scaled index $C_{K',K}(E)$ just applying
an additive zero point equal to $-2.5\log_{10}(1700/683)$. We find the happy
coincidence that this zero point is almost exactly equal to $-1$, in fact it is 
$-0.990$, allowing the immediate conversion from the normalized scotopic -- photopic index to
its scaled version:

\begin{center}
\begin{tabular}{lrr}
\hline
\multicolumn{1}{c}{$E$} & 
\multicolumn{1}{c}{$C_{K',K}(E)$} &
\multicolumn{1}{c}{$S/P$} \\
\hline
PC amber     &    1.042 & 0.383 \\
Metal halide &   -0.469 & 1.540 \\
Incandescent &   -0.194 & 1.196 \\
\hline
\end{tabular}
\end{center}

In the scaled version $C_{K',K}(E)$, a null value would mean ''same efficacy (or action) 
in both bands''. The figures indicate that the PC amber LED exerts an action some 2.5 times
more intense photopically than scotopically, but the contrary happens
with the other two lamps. The incandescent one is closer to the equilibrium
of actions (null value), while the metal halide lamp displays a clearly stronger
scotopic action. The classical $S/P$ ratio can be derived from the scaled
index through Eq. \ref{C_t_Q}.

\begin{figure}
\resizebox{\textwidth}{!}{\includegraphics {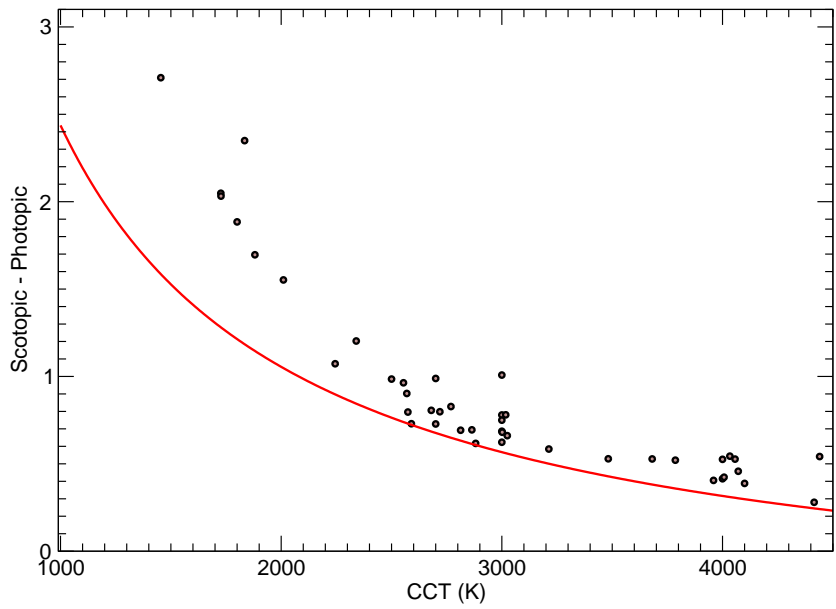}}
\caption{The scotopic -- photopic normalised ($V'-V$) spectral index (a measure
of the $S/P$ ratio) as a function of CCT, for the set of lamps discussed
in Sect. \ref{sample}. The correlation is poor, implying that
CCT is not a good indicator for $S/P$ considerations. The solid line marks the 
location of black bodies.}\label{SP}
\end{figure}

CCT is a poor proxy of the $S/P$ ratio, as can be seen in Fig. \ref{SP},
that shows the normalized index $C_{V',V}$ as a function of CCT, for the 
set of lamps discussed in Sect. \ref{sample}.

Indices related to the $S/P$ ratio may be of specific interest in lighting, but they
are not specially useful to classify lamps according to their content of
blue light, because the two filters $V'$ and $V$ are very close to each other,
and they show a significant overlap. Studies on the environmental and health effects
of artificial light at night place focus on the limitation of blue light and,
for this purpose,
}
at some point, some
small set of standard filter pairs should arise from a consensus
among the scientific and technical community interested
in lamp characterization. Without any intention of making
a firm proposal in this sense, and with the only aim of
illustrating the system, we will consider the spectral index
defined by means of the melanopic curve $Z$ (Fig.~\ref{CIEcurves}, Table
\ref{Johnson_tab}) and the photopic $V$ function. The melanopic curve,
with effective wavelength around $\lambda = 495$ nm, 
is related to the pigment active in the { intrinsically photosensitive retinal
ganglion cells}, and it is linked to the regulation of the
human circadian system, a matter of special interest for studies
on artificial light at night and chronodisruption. The resulting 
$Z-V$ index may act as a measure of the input to the ganglion
cells per each unit of photopically useful light. Often, 
redder (larger) values of this index will be preferred,
meaning a smaller amount of potentially disruptive light 
(from a circadian point of view) per lumen.

The ancillary data in \cite{gal17} give this index for a large set of lamps. Here we show the
values that are obtained for the three examples that we have
been using in previous sections:

\begin{center}
\begin{tabular}{lrr}
\hline
\multicolumn{1}{c}{$E$} & 
\multicolumn{1}{c}{$C_{Z,V}(E)$} &
\multicolumn{1}{c}{$C_{Z,V}(E)$} \\
\hline
PC amber      &  2.937 & 0.067 \\
Metal halide  &  0.787 & 0.484 \\
Incandescent  &  1.082 & 0.369 \\
\hline
\end{tabular}
\end{center}

{ Finally, let us comment that the Chilean regulation for their
northern astronomical regions, \cite{chile13}, specifies 
limits both on the amount of blue, and on the 
amount of red emission, compared to the quantity of light emitted
in the central part of the visible interval. We express these criteria
as conditions on indices built from window filters, in Table 
\ref{legal}.}

\section{Application to a lamp sample}
\label{sample}

We illustrate the formalism deriving a set of selected
spectral indices for a sample of more than sixty lamp spectra.
The contents of the database are described in detail in
\ref{contents}, and they are available at \cite{gal17}. In this section 
we discuss some conclusions that can be drawn from those results, and 
we compare our spectral index proposal with several similar ideas found
in the literature. 

\subsection{Discussion}
\label{discussion}
The results that can be found in the ancillary data set, \cite{gal17}, lead to several
conclusions. 

\begin{figure}
\resizebox{\textwidth}{!}{\includegraphics {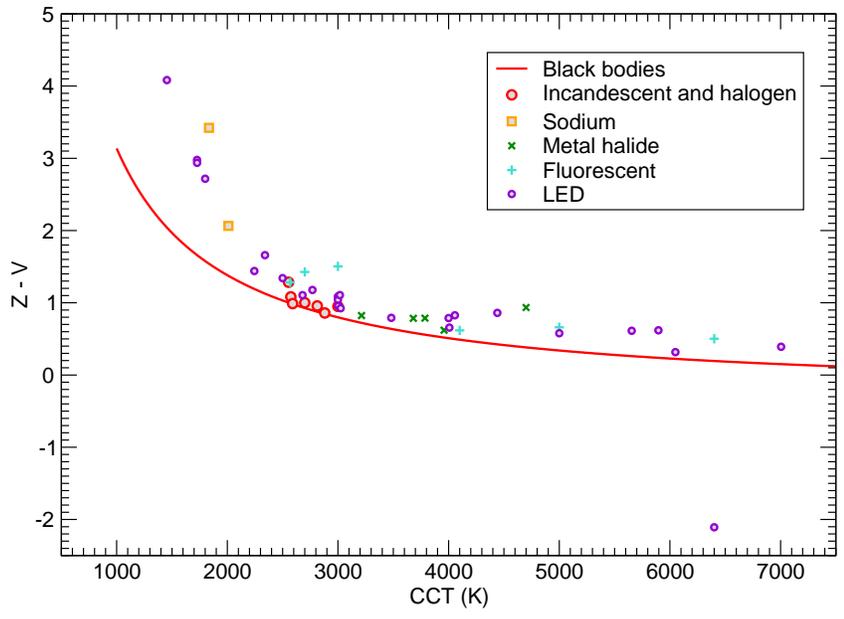}}
\caption{The melanopic -- photopic ($Z-V$) spectral index (a measure
of the circadian input per lumen) as a function of CCT. }\label{ZVCCT}
\end{figure}

The $Z-V$ melanopic -- photopic index provides some insight into the {\it blueness} of light sources,
in relation to the amount of useful light emitted. Fig.~\ref{ZVCCT} displays
the relation between this index and CCT for the whole sample of light sources studied. 
The solid line represents the black body locus. We see that, in general, all 
artificial lamps are redder than black bodies with the same CCT (spectral indices
are larger than those of black bodies), a tendency that is specially
strong towards low CCT values. The LED lamp with negative $Z-V$ index 
(source number 60) is
a very special device used for signaling, not for lighting purposes. We
clearly see a general correspondence between CCT and $Z-V$, but it is evident
that this relation shows a significant spread, a spread that intensifies, too,
in the low CCT area. 

\begin{figure}
\resizebox{\textwidth}{!}{\includegraphics {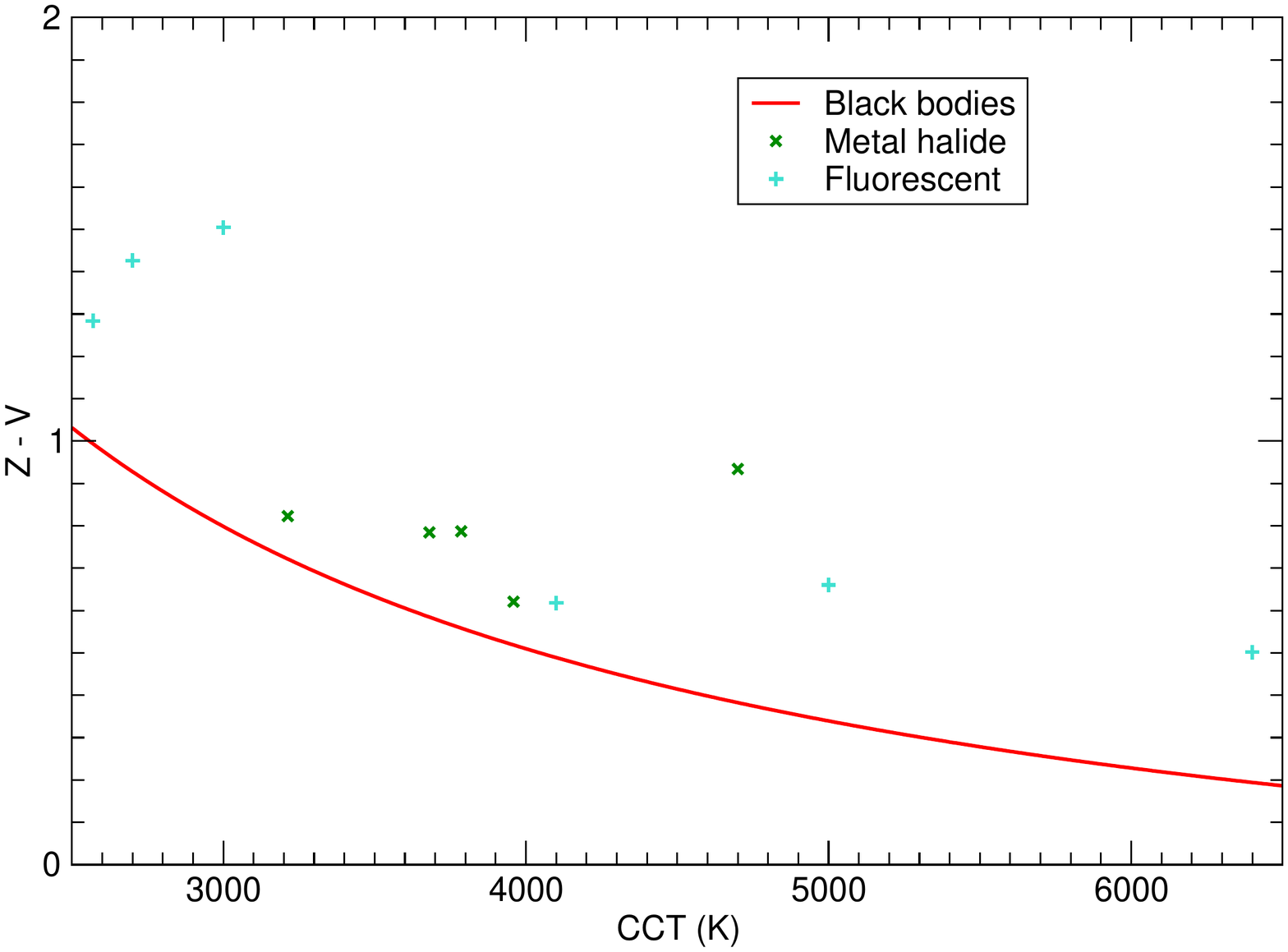}}
\caption{The melanopic -- photopic ($Z-V$) spectral index as a function of CCT
for fluorescent and metal halide lamps. The lack of correlation is specially
outstanding for these technologies.} \label{flu_hal}
\end{figure}

The correlation with CCT is worse for those sources whose spectra are more 
different to those of black bodies. Fig.~\ref{flu_hal} shows this for
discharge lamps of two different technologies: 
fluorescent (low pressure mercury) and metal halide. Let us underline
that in the second case, the lamp with highest CCT (number 27) is the reddest for this
filter pair, contrary to what may be expected if CCT would be a good spectral descriptor. 

\begin{figure}
\resizebox{\textwidth}{!}{\includegraphics {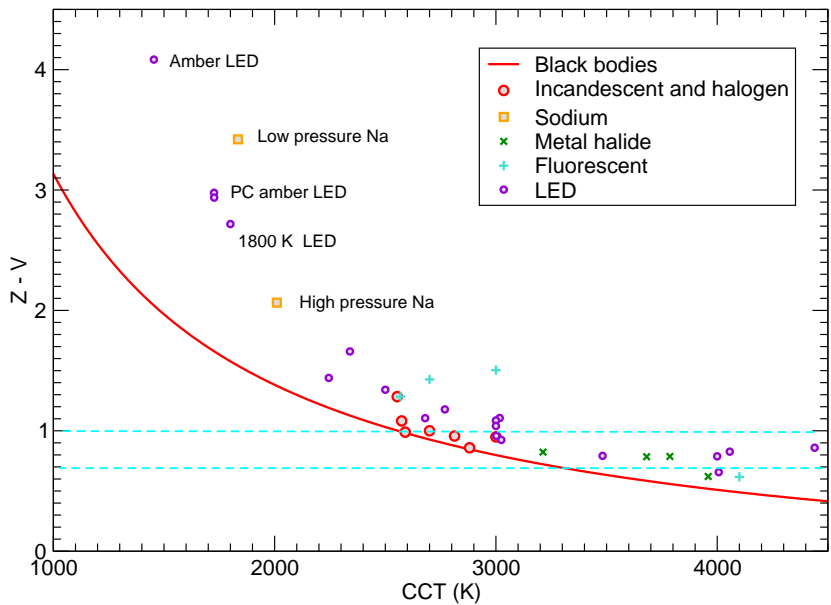}}
\caption{The melanopic -- photopic ($Z-V$) spectral index as a function of CCT
for sources below CCT = 4500 K. Inspecting this graph illustrates what kind
of mistakes may be possible when classifying lamps according to CCT as 
if it was a measure of the amount of blue light in the spectra, even when
dealing with only one technology. (For interpretation of the references to color in this figure legend, the reader is referred to the web version of this article.)} \label{lowCCT}
\end{figure}

Zooming into the low CCT area (Fig.~\ref{lowCCT}), we can see that the spread of the relation
makes CCT almost meaningless as a descriptor of {\it blueness} for this pair of 
filters, for CCT values larger than 2500 K, approximately. Over an interval that 
covers a span of some 2000 K, the spectral index $Z-V$ varies from 
0.7 to 1.0, only 0.3 magnitudes, and not always in a monotonic way.
LEDs clustering around CCT = 3000 K show a similar variation
in $Z-V$, from 0.9 to 1.2 magnitudes.

The region of the extremely low CCT lamps (below 2500 K, the so-called
'warm-light emitters') shows the largest spread. Non-standard LEDs
clearly demonstrate their potential as light emitters with blue-light
content levels even lower than those of the traditional sodium discharge lamps. 
However, let us keep in mind that such devices are currently seldom used actually for 
real lighting, where much bluer LEDs are normally used, in the CCT
interval over 2500 K, usually with much bluer $Z-V$ values, around or
below $Z-V=1$. 

\begin{figure}
\resizebox{\textwidth}{!}{\includegraphics {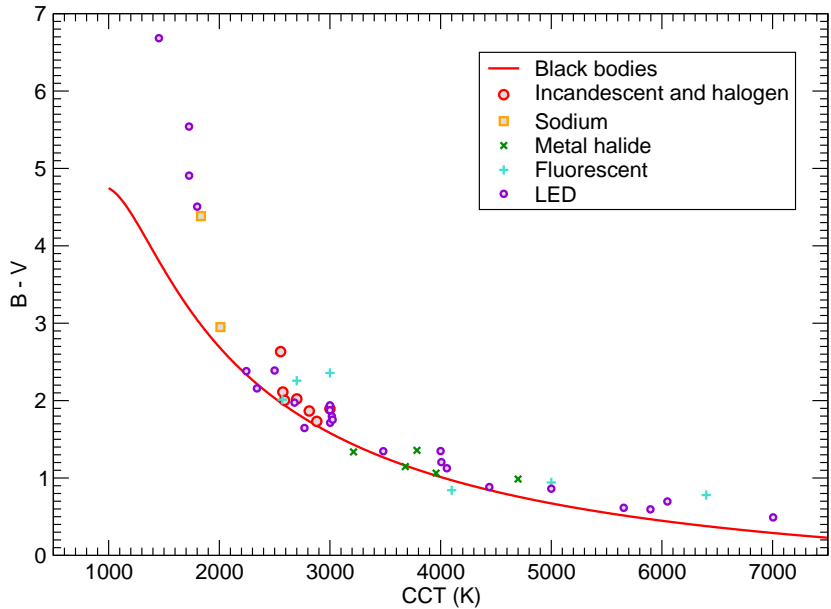}}
\caption{The blue hazard -- photopic ($B-V$) spectral index as a function of CCT. (For interpretation of the references to color in this figure legend, the reader is referred to the web version of this article.)} \label{bluehazardCCT}
\end{figure}

Index $B-V$ (blue-hazard -- photopic) shows a better correlation with CCT, 
but still with a significant scatter. Redder light sources (CCT $<$ 2000 K)
show an extreme 
deviation from theoretical black bodies (Fig.~\ref{bluehazardCCT}). Several lamps
are slightly bluer than the corresponding black bodies for this filter pair. 
In this index, non-standard LEDs are even redder than low-pressure sodium 
(pure amber LED, PC amber LED, Philips 1800 K LED). 

As commented in Sect.~\ref{curves}, attention should be paid to
the results derived for the aphakic--photopic index, due to
the non-standard character of the aphakic $A$ curve, { displayed
in Fig. \ref{blue_hazard}.
The $A$ curve is an adaptation of the blue-hazard $B$ curve, to take
into account the increased transparency of eye tissues to blue
light in very young persons. It cannot be normalized to maximum
unity because it has to follow the profile of curve $B$ at the red side.
For this reason, aphakic indices are always much bluer (smaller) than the 
equivalents built from the standard blue-hazard function. From the pair of
indices given in the ancillary data set, $B-V$ and $A-V$, a third one can be very easily
deduced as the difference: $A-B$. This index will always be negative, and it provides
a measure of how the potential risk of the blue light from a lamp
may be increased for very young subjects.}

{
We have already commented the similarity among the astronomical
filter $V_{\mbox{\footnotesize J}}$ and the photopic function $V$.
The blue-hazard $B$ curve is not too far from the Johnson-Cousins
$B_{\mbox {\footnotesize J}}$ filter. As a result, there exists a 
significant correlation between the values of the indices
$B-V$ and $B_{\mbox {\footnotesize J}}-V_{\mbox{\footnotesize J}}$,
but it is not good enough to forget the differences. So, care has to be
taken not to confuse the perceptual functions $B$ and $V$ with their 
Johnson-Cousins astronomical close relatives, named $B_{\mbox{\footnotesize J}}$ 
and $V_{\mbox{\footnotesize J}}$ in this paper, but labelled exactly with
the same symbols $B$ and $V$ in the astronomical literature.}

\begin{figure}
\resizebox{\textwidth}{!}{\includegraphics {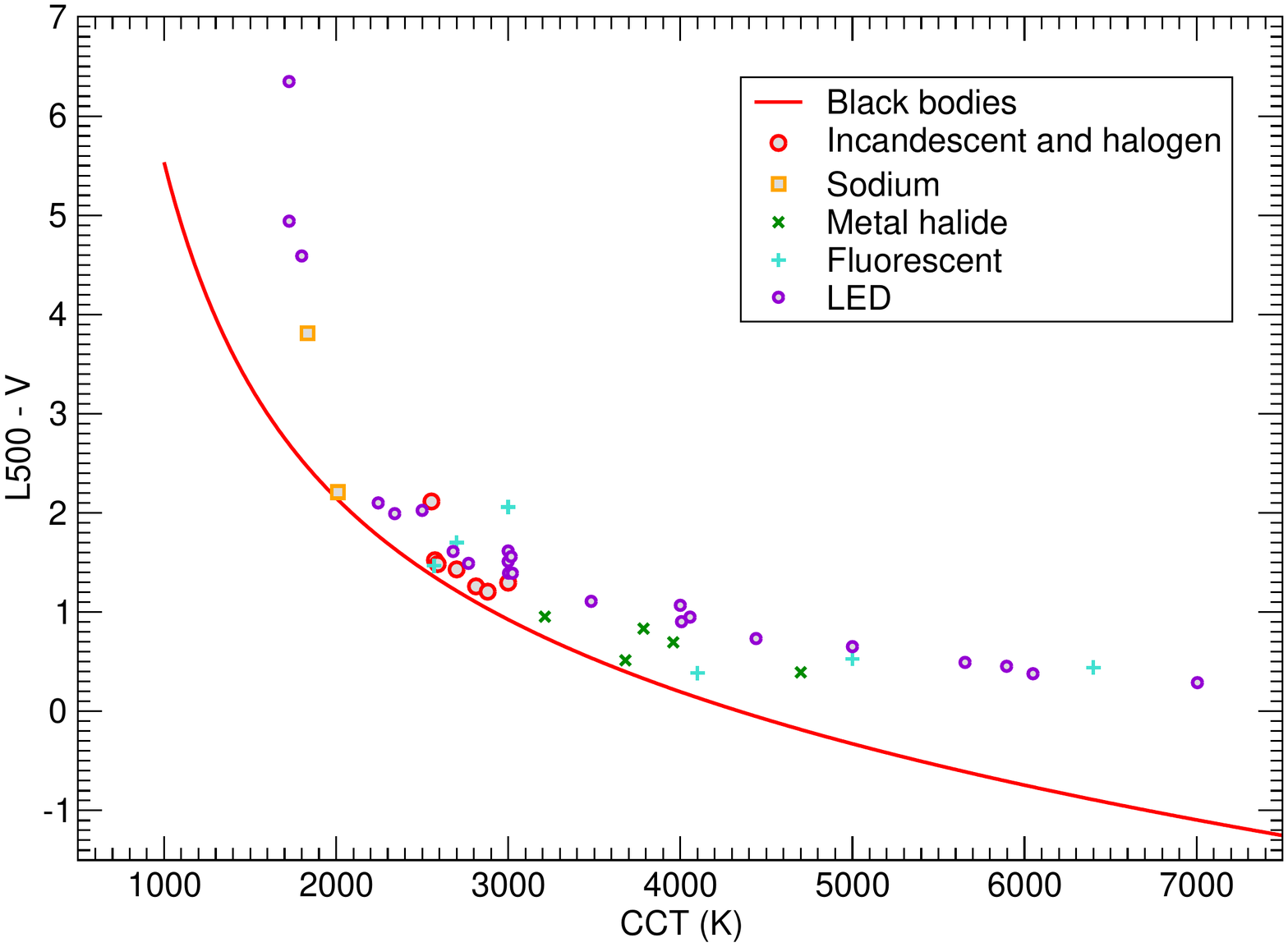}}
\caption{The $L_{500}$ -- photopic ($L_{500}-V$) spectral index as a function of CCT.} \label{L500CCT}
\end{figure}

Index $L_{500}-V$ has some chances to become a kind of standard to
classify lamps according to their amount of blue emission, if we want to 
evaluate this per unit of photopically efficient light (let us say, per lumen). 
{ Regulations already in use in Canary Islands (\cite{iac17}, Table \ref{legal})
rely on this kind of criteria.}
As we see in Fig.~\ref{L500CCT}, CCT would be a bad
descriptor for the evaluation of the content of blue light, 
{\it specially in the interval of the 
most frequent CCTs}, from 2000 to 4000 K. All lamps analyzed are redder than
the black bodies with the same CCT. 

\subsection{Comparison with alternative formalisms}
\label{alternative}
The specific { features} of lamp spectra characterization by means of the 
spectral index system can be enumerated as follows:
\begin{enumerate} 
\item The system is quantitative, providing the numerical and physical meaning
that CCT lacks.
\item Flexibility and general character, free election of filter pairs
according to the needs of each field of study. 
\item Unit-independent. Any index can be computed from any spectrum, expressed
in the units you want. The mandatory comparison of two bands extracted from the
same spectrum eliminates any worry about units. We may say that the 
index is self-normalized.
\item As a consequence of the previous point, no reference source
has to be defined or used. Each spectrum acts as its own calibrator.
\item Standard data already obtained at labs for lamp characterization
are perfect for spectral index computation. The calculations are even 
simpler than those routinely performed for CCT determination. 
\item The logarithmic nature allows lamp comparison through 
simple additions and subtractions, { as well as the inclusion of
scale factors in the form of additive zero points}.
\item The formalism is fully compatible with that already in use in
astronomy, specially in the field of studies on artificial sky brightness and
light pollution. 
\end{enumerate}

A spectral index specifies an elemental physical output from
lamps, at a very low level. For instance, when using pigment sensitivity curves
as those showed in Fig.~\ref{CIEcurves}, the resulting indices would describe
the direct physical-chemical {\it input} on the corresponding 
sensitive cells. The possible relation of this elementary input to
more complex physiological effects falls out of the scope of the formalism.
We may, again, pose a comparison with astrophysics, where stellar spectra are
described very often by means of the Johnson index 
$B_{\mbox{\footnotesize J}}-V_{\mbox{\footnotesize J}}$ (do not confuse with 
our blue hazard -- photopic index defined above). That index constitutes an
elementary description of the star, and establishing any further
relation between this index and more complex quantities, such as
effective temperature, falls on the side of the applications of the 
measurement, not on the side of performing the measurement itself.

Of course, the idea of describing lamp spectra performing computations 
from the integration on several bands has been present in lighting 
engineering from the beginning, and definitions such as that of luminous efficacy 
{ or $S/P$ ratio}
are already based on concepts of this kind. In recent years, with the 
increasing need of going multi-band, several proposals have circulated, pointing to ideas
quite close to the spectral index formalism described in this paper. We will briefly comment
a representative subset of them.

\cite{zvp12} study the way to
optimize solid-state lamps for photobiologically friendly
lighting. This leads them to consider quotients of integrated fluxes defined in 
a fashion similar to what we state in Eq.~\ref{int_flux}. The main
differences are that they apply { scaling}  factors (in units of
lm/W), and that their second filter is always the bolometric one, i.e., they
normalize their 'luminous efficacy of radiation' estimators
(equations 1, 2 and 3 of their paper) according to the total integrated flux
emitted by the lamps. Their 'circadian efficacy of radiation' estimator is
very similar to our quotient of integrated fluxes $\Phi_{Z}(E)$/$\Phi_{\mbox{\footnotesize bol}}$.
The quotients of such estimators expressed in their equation 6
would be equivalent to a {\it difference} of our spectral indices (their dimensional
multiplicative factors would transform into just an additive,
spectrum-independent, zero point, by the way). 
Later on in the paper, the authors introduce these elementary
quantitative estimators inside a non trivial model, to
derive figures of merit for their scientific purposes, a process that
may have been done using spectral indices too.

\cite{aub13} developed a system that received even a denomination
similar to ours. They define quotients of
integrated fluxes very close to our $\Phi_{1}(E)$/$\Phi_{V}(E)$; note that this
time the second filter is always the photopic one, in what they call 'constant 
lumen normalization' (equation 3 in their paper). After defining three filters
of interest to be used as $F_1$, they compute their final 'indices' 
relative to the values of a standard illuminant (specifically the 
ICE standard illuminant D65). In the language of our system, this would translate
into applying the corresponding spectral index of the standard illuminant
as a subtractive zero point, due to the logarithmic nature of our proposal. 
For instance, in our scheme, $C_{Z,V}(D65)=0.186$ and, thus, the same
index for any other spectrum may be referred to the D65 scheme just subtracting
this value: then, $C_{Z,V}(E)-C_{Z,V}(D65)=0$ would mean 'the same $Z/V$ ratio
as the standard illuminant', while in our simpler scheme (not relative to any standard illuminant)
$C_{Z,V}(E)=0$ has a meaning closer to the physical reality: 
'same energy in both filters'.

The sound work by \cite{aub13} is not totally general. 
It is doubtful whether it is really necessary to rely on a standard illuminant as reference.
The non-logarithmic character places the work further from astronomical 
tradition (what may be more relevant for their 'star light index') and, while
turning easier the determination of their 'indices', it makes somewhat more cumbersome
their later management in practical use. Their specific selection of filters is just one among 
many other possible, but maybe they are too complicated, mixing 
simple physical inputs with non-trivial considerations about effects and
actions that, in our opinion, should be left for later stages in the interpretation.
We pursue the computation of simple numbers as close as possible to the true, 
native and neat properties of the spectra.

Finally, \cite{eyb15} delve into the complexities of circadian 
inputs going back to the elementary concept of integrated filtered flux
of this work (Eq.~\ref{int_flux}), and of \cite{zvp12}, to later combine
several filters (or 'weighting functions') in a shape that may have been 
formulated, too, in our language of spectral indices. { In our opinion, \cite{eyb15}
offer an interesting example of clean separation of inputs and actions or effects,
with an approach at only one step from using a completely general formalism for the multiplicity
of filters used by them. These authors deal with those filters on a one-by-one basis,
handling a complex network of scaling factors and standards that we avoid in our spectral 
index system, seeking maximum simplicity and homogeneity. }
\newline \newline \newline
\noindent\textit{Acknowledgments}

\noindent The computations have been performed on lamp spectra kindly provided,
for the exclusive purposes of this work, by: Manuel Garc\'{\i}a Gil (Generalitat de 
Catalunya, Servei per a la Prevenci\'o de la Contaminaci\'o Lum\'{\i}nica), Mar Gandolfo 
de Luque (Comit\'e Espa\~nol de Iluminaci\'on and Philips Spain), Javier D\'{\i}az de Castro 
(Instituto de Astrof\'{\i}sica de Canarias), Laura Guzm\'an Varo (Comit\'e Espa\~nol de 
Iluminaci\'on and Light Environment Control), and Ramon Llorens (SACOPA-IgniaLight). 
We have also used spectra downloaded from the public database LSPDD: Light Spectral Power Distribution 
Database (www.lspdd.com/). The author is very thankful to all of them. Without this 
excellent base material, this work would not have been possible. \newline \newline
{\bf Funding:} This research did not receive any specific grant from funding agencies 
in the public, commercial, or not-for-profit sectors. \newline\newline
{\bf Conflicts of interest:} none.\newline\newline

\appendix
\section{Lamp sample: contents of the database}
\label{contents}

The spectral index formalism has been applied to a database with more
than sixty spectra kindly provided by the sources mentioned in the acknowledgments
section, for the only purposes of this work. We have also used a set of lamp spectra 
from the { public open repository Light Spectral Power Distribution Database,  \cite{lspdd17}. }
Several black body spectra have been generated by us in a straightforward way. 
The results are available at \cite{gal17}, and they cover different 
lighting technologies, including experimental spectra obtained with different spectrometers. 
For each available spectrum we compute the following spectral indices:

\begin{itemize}
\item \textbf{Photopic -- bolometric}, $V-$bol. A measure of 
the luminous efficacy of lamps, showing the fraction of spectral 
energy emitted inside the photopic band. For this index, bluer lamps
(with lower values of the spectral index) would be more efficient from a lighting
point of view. 
\item \textbf{Lowpass-$\lambda$ 500 nm -- photopic}, $L_{500}-V$. Comparing
the amount of energy emitted below 500 nm with that efficient for
lighting purposes, in photopic conditions. A good measure of the
amount of blue light compared to lighting efficacy. { If the aim is
to reduce the amount of blue light, then larger values of this index are preferred.}
\item \textbf{Melanopic -- photopic}, $Z-V$. It compares the amount of
light active on the circadian receptors with the intensity efficient for lighting 
in photopic conditions. { Again, larger values are better, in the sense that
they indicate a lower input on the intrinsically sensitive retinal ganglion cells.}
\item \textbf{Lowpass-$\lambda$ 500 nm -- bolometric}, $L_{500}-$bol. Reflects the 
ratio between blue light and the total amount of energy radiated by the lamp.
\item \textbf{Blue hazard -- photopic}, $B-V$, and \textbf{aphakic -- photopic},
$A-V$, according to the curves specified by \cite{icnirp13}. 
\end{itemize}

The results are shown for all lamps in the ancillary data attached to this 
article. For each index, the value $C_{1,2}(E)$ according to definition 
\ref{spectral_index} is given but, also, for a better understanding
of the system for people not well acquainted with the logarithmic scale,
the relative integrated flux $Q_{1,2}(E)$ (Eq.~\ref{rel_flux}) is provided. A last column
contains the value of the correlated color temperature (CCT) of the lamps, 
in kelvins, directly drawn from the sources that provided the spectra.

A graphic annex displays the spectra for 61 of the 69 light sources.




\end{document}